\documentclass[12pt,preprint]{aastex}
\usepackage{natbib}
\input epsf

\newcount\longrefs
\def\aap{\ifnum\longrefs=1 {Astron.\ Astrophys.}\else 
                           {A\hbox{\rm \&}A}\fi}
\def\aapr{\ifnum\longrefs=1 {Astron.\ Astrophys.\ Rev.}\else 
                            {A\hbox{\rm \&}AR}\fi}
\def\aaps{\ifnum\longrefs=1 {Astron.\ Astrophys.\ Suppl.}\else 
                            {A\hbox{\rm \&}AS}\fi}
\def\aj{\ifnum\longrefs=1 {Astron.\ J.}\else 
                          {AJ}\fi} 
\def\ao{\ifnum\longrefs=1 {Applied Optics}\else 
                           {Appl.\ Opt.}\fi} 
\def\aspcs{\ifnum\longrefs=1 {Astron.\ Soc.\ Pacific Conf. Series}\else 
                           {ASP Conf.\ Ser.}\fi} 
\def\apj{\ifnum\longrefs=1 {Astrophys.\ J.}\else 
                           {ApJ}\fi} 
\def\apjl{\ifnum\longrefs=1 {Astrophys.\ J. Lett.}\else 
                            {ApJ}\fi} 
\def\aplett{\ifnum\longrefs=1 {Astrophys.\ J. Lett.}\else 
                            {ApJ}\fi} 
\def\apjs{\ifnum\longrefs=1 {Astrophys.\ J. Suppl.}\else 
                            {ApJS}\fi}
\def\apss{\ifnum\longrefs=1 {Astrophys.\ and Space Science}\else 
                            {Ap\hbox{\rm \&}SS}\fi}
\def\araa{\ifnum\longrefs=1 {Ann.\ Rev.\ Astron.\ Astrophys.}\else 
                            {ARA\hbox{\rm \&}A}\fi}
\def\azh{\ifnum\longrefs=1 {Astronomicheskii Zhurnal}\else 
                            {Astron.\ Zhur.}\fi}
\def\baas{\ifnum\longrefs=1 {Bull.\ Am.\ Astron.\ Soc.}\else 
                            {BAAS}\fi}
\def\bain{\ifnum\longrefs=1 {Bull.\ Astronom.\ Institutes Netherlands}\else
                            {Bull.\ Astr.\ Inst.\ Neth.}\fi}
\def\gca{\ifnum\longrefs=1 {Geochim.\ Cosmochim.\ Acta}\else 
                           {Geochim.\ Cosmochim.\ Acta}\fi}
\def\grl{\ifnum\longrefs=1 {Geophys.\ Res.\ Lett.}\else 
                           {Geoph.\ Res.\ Lett.}\fi}
\def\iaucirc{\ifnum\longrefs=1 {IAU Circulars}\else 
                          {IAU Circ.}\fi}
\def\ip{\ifnum\longrefs=1 {in press}\else 
                          {in press}\fi}
\def\jchemp{\ifnum\longrefs=1 {J.\ Chem.\ Phys.}\else 
                           {J.\ Chem.\ Phys.}\fi}  
\def\jcp{\ifnum\longrefs=1 {J.\ Chem.\ Phys.}\else 
                           {J.\ Chem.\ Phys.}\fi}  
\def\jgr{\ifnum\longrefs=1 {J.\ Geophys.\ Res.}\else 
                           {J.\ Geophys.\ Res.}\fi}  
\def\jmolspec{\ifnum\longrefs=1 {J.\ Mol.\ Spectrosc.}\else 
                           {J.\ Mol.\ Spectrosc.}\fi}  
\def\jqsrt{\ifnum\longrefs=1 {J.\ Quant.\ Spectrosc.\ Radiat.\ Transfer}\else 
                           {J.\ Quant.\ Spectrosc.\ Radiat.\ Transfer}\fi}  
\def\jrasc{\ifnum\longrefs=1 {J.\ Royal Astron.\ Soc.\ Canada}\else 
                           {JRAS Can.}\fi}  
\def\mnras{\ifnum\longrefs=1 {Mon.\ Not.\ Roy.\ Astron.\ Soc.}\else 
                             {MNRAS}\fi} 
\def\nat{\ifnum\longrefs=1 {Nature}\else 
                           {Nat}\fi}
\def\pasj{\ifnum\longrefs=1 {Pub.\ Astron.\ Soc.\ Japan}\else 
                            {PASJ}\fi} 
\def\pasp{\ifnum\longrefs=1 {Pub.\ Astron.\ Soc.\ Pacific}\else 
                            {PASP}\fi} 
\def\physscr{\ifnum\longrefs=1 {Physica Scripta}\else 
                            {Phys.\ Scrip.}\fi} 
\def\planss{\ifnum\longrefs=1 {Planetary \& Space Science}\else 
                            {Plan. \& Space Sci.}\fi} 
\def\procspie{\ifnum\longrefs=1 {Proc.\ SPIE}\else 
                            {Proc.\ SPIE}\fi} 
\def\qjras{\ifnum\longrefs=1 {Quarterly J.\ Royal Astron.\ Soc.}\else 
                            {QJRAS}\fi} 
\def\sa{\ifnum\longrefs=1 {Soviet Astron..}\else 
                               {Sov.\ Astron.}\fi}
\def\skytel{\ifnum\longrefs=1 {Sky \& Telescope}\else 
                            {Sky \& Tel.}\fi} 
\def\solphys{\ifnum\longrefs=1 {Solar Phys.}\else 
                               {Solar Phys.}\fi}
\def\ssr{\ifnum\longrefs=1 {Space Science Rev.}\else 
                               {Space\ Sci.\ Rev.}\fi}

\def\nl{,\ } %%\def\nl{\newline}  %% redefine as \newline for mail addresses

\def\dutch{\def\refname{Referenties}\def\abstractname{Samenvatting}%
  \def\bibname{Bibliografie}\def\chaptername{Hoofdstuk}%
  \def\appendixname{Bijlage}\def\contentsname{Inhoudsopgave}%
  \def\listfigurename{Lijst van figuren}\def\listtablename{Lijst van tabellen}%
  \def\indexname{Index}\def\figurename{Figuur}\def\tablename{Tabel}%
  \def\partname{Deel}\def\enclname{Bijlage(n)}\def\ccname{Ter attentie van}%
  \def\headtoname{Aan}\def\headpagename{Pagina}%
  \def\today{\number\day\space\ifcase\month\or januari\or februari\or maart\or%
     april\or mei\or juni\or juli\or augustus\or september\or oktober\or%
     november\or december\fi \space\number\year}%
  \typeout{
              >>>>> use hlatex209 for Dutch hyphenation <<<<< 
         }}
\newcounter{onefig} \newcounter{fignumber}
\newcount\nocaptions \newcount\nofigures \newcount\figwidth
\newcount\viewgraphs
  \def\paper{}  \def\figlabel{} 
\long\def\nextfig#1{\setcounter{figure}{\value{fignumber}}
  \addtocounter{fignumber}{1}
  \ifnum \viewgraphs=1 \newpage \pagestyle{empty} \fi 
  \ifnum\value{onefig}=0 #1 \fi                 
  \ifnum\value{onefig}=\value{fignumber} #1 \fi}
\def\figwidths#1#2{\ifnum \nocaptions=1 #2mm \else #1mm \fi}  
\def\paper#1{}  %% redefine for separate-figure identification line
\long\def\plotfig#1#2{\ifnum \nofigures=1 \else #2 \fi}
\long\def\captiontext#1{\ifnum \nofigures=1 \raggedright \fi 
   \ifnum \nocaptions=1 \paper
     \ifnum \viewgraphs=0 
       \newline  \mbox{}\hrulefill\mbox{} \newline 
       \newline label:~\{\figlabel\} 
     \fi 
     \else \ifnum \nofigures=0 \fi 
   #1 \fi}

\newcount\panelwidth \newcount\panelheight 
\newcount\bxmin \newcount\bymin \newcount\bxmax \newcount\bymax
\newcount\tbxmin \newcount\tbymin
\newcount\tpanelwidth \newcount\tpanelheight \newcount\tpdif
\def\panelsize #1,#2;{\panelwidth=#1 \panelheight=#2}  
\def\setbb #1,#2;#3,#4;#5,#6;{% UNITS: bp (from ghostview)
  \tbxmin=#1 \tbymin=#2    %% full box (axis titles) lower left corner
  \bxmin=#3 \bymin=#4      %% bare box (ticks only) lower left corner
  \bxmax=#5 \bymax=#6}     %% upper right corner
\def\barepanel #1{%
  \ifnum\panelheight=0 
    \tpdif=\bymax \advance\tpdif by -\bymin
    \multiply \tpdif by \panelwidth
    \tpanelheight=\tpdif
    \tpdif=\bxmax \advance\tpdif by -\bxmin
    \divide \tpanelheight by \tpdif
  \else \tpanelheight=\panelheight \fi
  \epsfig{file=#1,%
     bbllx=\bxmin bp,bblly=\bymin bp,bburx=\bxmax bp,bbury=\bymax bp,clip=,%
     width=\panelwidth mm,height=\tpanelheight mm}}
\def\labelypanel #1{% TeX permits only integer arithmetic, so bp and mm
  \ifnum\panelheight=0 
    \tpdif=\bymax \advance\tpdif by -\bymin
    \multiply \tpdif by \panelwidth
    \tpanelheight=\tpdif
    \tpdif=\bxmax \advance\tpdif by -\bxmin
    \divide \tpanelheight by \tpdif
  \else \tpanelheight=\panelheight \fi
  \tpdif=\bxmax \advance\tpdif by -\tbxmin
  \tpanelwidth=\panelwidth \multiply \tpanelwidth by \tpdif
  \tpdif=\bxmax \advance\tpdif by -\bxmin
  \divide \tpanelwidth by \tpdif
  \epsfig{file=#1,%
    bbllx=\tbxmin bp,bblly=\bymin bp,bburx=\bxmax bp,bbury=\bymax bp,%
    clip=,width=\tpanelwidth mm,height=\tpanelheight mm}}
\def\labelxpanel #1{%
  \ifnum\panelheight=0 
    \tpdif=\bymax \advance\tpdif by -\bymin
    \multiply \tpdif by \panelwidth
    \tpanelheight=\tpdif
    \tpdif=\bxmax \advance\tpdif by -\bxmin
    \divide \tpanelheight by \tpdif
  \else \tpanelheight=\panelheight \fi
  \tpdif=\bymax \advance\tpdif by -\tbymin
  \multiply \tpanelheight by \tpdif
  \tpdif=\bymax \advance\tpdif by -\bymin
  \divide \tpanelheight by \tpdif
  \epsfig{file=#1,%
    bbllx=\bxmin bp,bblly=\tbymin bp,bburx=\bxmax bp,bbury=\bymax bp,%
    clip=,width=\panelwidth mm,height=\tpanelheight mm}}
\def\labelxypanel #1{%
  \ifnum\panelheight=0 
    \tpdif=\bymax \advance\tpdif by -\bymin
    \multiply \tpdif by \panelwidth
    \tpanelheight=\tpdif
    \tpdif=\bxmax \advance\tpdif by -\bxmin
    \divide \tpanelheight by \tpdif
  \else \tpanelheight=\panelheight \fi
  \tpdif=\bxmax \advance\tpdif by -\tbxmin
  \tpanelwidth=\panelwidth \multiply \tpanelwidth by \tpdif
  \tpdif=\bxmax \advance\tpdif by -\bxmin
  \divide \tpanelwidth by \tpdif 
  \tpdif=\bymax \advance\tpdif by -\tbymin 
  \multiply \tpanelheight by \tpdif
  \tpdif=\bymax \advance\tpdif by -\bymin
  \divide \tpanelheight by \tpdif
  \epsfig{file=#1,%
    bbllx=\tbxmin bp,bblly=\tbymin bp,bburx=\bxmax bp,bbury=\bymax bp,%
    clip=,width=\tpanelwidth mm,height=\tpanelheight mm}}

\def\CC{\par \vspace*{-2ex} \footnotesize \baselineskip=8pt \begin{verbatim}}

\long\def\startignore #1\stopignore{}   %% use \startignore....\stopignore

\def\setlistparams{         
  \topsep=0.7ex                 %% ADAPT: parskip=0: 0.7;  parskip=1: -1.2ex
  \itemsep=0.7ex                %% space between items
  \leftmargini=3ex}             %% dashes at beginning of line 
\setlistparams                  %% recall after type changes 

\newcounter{alistindex}       %% problems: a)  b) etc

\newcounter{romenumnr}

\newlength{\minipagewidth}

\newsavebox{\boxcontent}
\newcommand{\ovalhead}[1]{
  \unitlength=1cm
  \sbox{\boxcontent}{\mbox{~~{#1}~~}}
  \begin{center}
    \ifdim\wd\boxcontent>6ex 
    \ifdim\wd\boxcontent<8cm 
    \begin{picture}(8,3) \thicklines     
      \put(4.0,0.8){\oval(8,1.6)} 
      \put(0.0,0.7){\parbox{8cm}{
         \begin{center} \usebox{\boxcontent} \end{center}}}
    \end{picture}
    \else \ifdim\wd\boxcontent<12cm 
    \begin{picture}(12,3) \thicklines     
        \put(6.0,0.8){\oval(12,1.6)} 
        \put(0.0,0.7){\parbox{12cm}{
           \begin{center} \usebox{\boxcontent} \end{center}}}
    \end{picture}
    \else
    \begin{picture}(16,3) \thicklines     
        \put(8.0,0.8){\oval(16,1.6)} 
        \put(0.0,0.7){\parbox{16cm}{
           \begin{center} \usebox{\boxcontent} \end{center}}}
    \end{picture}
    \fi \fi \fi
  \end{center}} 

\newcounter{headnr}            
\newcounter{subheadnr}[headnr]
\newcounter{subsubheadnr}[subheadnr]
\def\head #1\par{
  \stepcounter{headnr}                          %% sets subheadnr = 0 too 
  \vspace{2ex} \noindent                        %% 2ex = space above, no *
  {\bf \theheadnr~~~~#1}\\[1ex] \noindent}      %% 1ex = space below
\def\subhead #1\par{  
  \stepcounter{subheadnr}
  \vspace{1.3ex} \noindent
  {\bf \theheadnr.\arabic{subheadnr}~~~#1}\\[0.3ex] \noindent}
\def\subsubhead #1\par{
  \stepcounter{subsubheadnr}
  \vspace{1.0ex} \noindent
  {\bf \theheadnr.\arabic{subheadnr}.\arabic{subsubheadnr}~~~#1}\\ \noindent}

\font\dropfont= cmr12 scaled \magstep5
\def\dropcap#1#2{{\noindent
    \setbox0\hbox{\dropfont #1}\setbox1\hbox{#2}\setbox2\hbox{(}%
    \count0=\ht0\advance\count0 by\dp0\count1\baselineskip
    \advance\count0 by-\ht1\advance\count0by\ht2
    \dimen1=.5ex\advance\count0by\dimen1\divide\count0 by\count1
    \advance\count0 by1\dimen0\wd0
    \advance\dimen0 by.25em\dimen1=\ht0\advance\dimen1 by-\ht1
    \global\hangindent\dimen0\global\hangafter-\count0
    \hskip-\dimen0\setbox0\hbox to\dimen0{\raise-\dimen1\box0\hss}%
    \dp0=0in\ht0=0in\box0}#2}

\def\level #1 #2#3#4{$#1 \: ^{#2} \mbox{#3} ^{#4}$}   

\def\sun{\hbox{$\odot$}}                   %% Sun symbol
\def\earth{\hbox{$\oplus$}}                %% Earth symbol

\def\mathstacksym#1#2#3#4#5{\def#1{\mathrel{\hbox to 0pt{\lower 
    #5\hbox{#3}\hss} \raise #4\hbox{#2}}}}

\mathstacksym\lta{$<$}{$\sim$}{1.5pt}{3.5pt} % less than approximately
\mathstacksym\gta{$>$}{$\sim$}{1.5pt}{3.5pt} % greater than approximately
\mathstacksym\lrarrow{$\leftarrow$}{$\rightarrow$}{2pt}{1pt} % equilibrium
\mathstacksym\lessgreat{$>$}{$<$}{3pt}{3pt} %% less or greater

\shorttitle{UV to mid-IR study of HD~164270 (WC9)}
\shortauthors{Crowther et al.}

\begin{document}
   \title{An Ultraviolet to Mid-Infrared Study of the Physical
and Wind Properties of HD~164270 (WC9) and Comparison to BD+30 3639 ([WC9])}

%\titlerunning{WC9 stars}

   \author{Paul A. Crowther}

   \affil{Department of Physics and
Astronomy, University of Sheffield, Hicks Building, Hounsfield Rd,
Shefffield, S3 7RH, UK}
   \email{Paul.Crowther@sheffield.ac.uk}

     \author{P. W. Morris}

\affil{NASA Herschel Science Center, Caltech M/S 100-22, 1200
E. California Blvd, Pasadena, CA 91125}

\author{J. D. Smith}

\affil{      Steward Observatory, University of Arizona, Tucson,
          AZ 85721}

%\author{          J. Houck }
%
%\affil{Astronomy Department, Cornell University, 106 Space Sciences Bldg.,
%Ithaca, NY 14853}
 
%\date{Received February 2005/Accepted}

   \begin{abstract} We present new {\em Spitzer} IRS observations of
HD~164270 (WC9, WR103). A quantitative analysis of the UV, 
optical, near- and mid-IR spectrum of HD~164270 is presented, allowing for 
line
blanketing and wind clumping, revealing $T_{\ast} \sim$48kK, $\log
L/L_{\odot} \sim$ 4.9, $\dot{M} \sim$ 10$^{-5} M_{\odot}$ yr$^{-1}$ for a
volume filling factor of $f \sim$0.1. Our models predict that He is
partially recombined in the outer stellar wind, such that recent
radio-derived mass-loss rates of WC9 stars have been underestimated. We
obtain C/He$\sim$0.2 and O/He$\sim$0.01 by number from optical
diagnostics. Mid-IR fine structure lines of [Ne\,{\sc ii}] 12.8 $\mu$m and
[S\,{\sc iii}] 18.7$\mu$m are observed, with [Ne\,{\sc iii}] 15.5$\mu$m
and [S\,{\sc iv}] 10.5$\mu$m absent. From these we obtain Ne/He $\sim$
Ne$^{+}$/He = 2.2$\times 10^{-3}$ by number, 7 times higher than the Solar
value (as recently derived by Asplund et al.), and S/He $\sim$S$^{2+}$/He
= 5.1$\times 10^{-5}$ by number. From a comparison with similar 
results
for other WC subtypes we conclude that WC9 stars are as chemically
advanced as earlier subtypes. We consider why late WC stars are
exclusively observed in high metallicity environments.  In addition, 
we
compare the UV/optical/mid-IR spectroscopic morphology of HD~164270 with
the Planetary Nebula central star BD+30 3639 ([WC9]). Their UV and optical
signatures are remarkably similar, such that our quantitative comparisons
confirm similarities in stellar temperature, wind densities and chemistry
first proposed by Smith \& Aller, in spite of completely different
evolutionary histories, with HD~164270 presently a factor of ten more
massive than BD+30 3639. At mid-IR wavelengths, the dust from the dense
young, nebula of BD+30 3639 completely dominates its appearance, in
contrast with HD~164270.
\end{abstract}


   \clearpage
%
%________________________________________________________________

\section{Introduction}

It has long been known that low ionization, carbon sequence Wolf-Rayet
stars (WC9 stars) are exclusively found within the inner, metal-rich
disk of our Milky Way  (van der Hucht 2001; Hopewell et al. 2005).
No WC9 stars are known elsewhere in the Local Group 
of galaxies, despite numerous surveys (Heydari-Malayeri et al. 1990, Moffat 1991). Indeed, WC9 populations have 
only recently been identified elsewhere within the metal-rich spiral 
M~83 by Hadfield et al. (2005).

The link between late WC subtypes and metal-rich environments led Smith \& 
Maeder (1991) to suggest that strong stellar winds at high metallicity reveal
He-burning processed material at an earlier phase than at lower metallicity,
which they attributed spectroscopically to late WC stars, i.e. a decreasing C/He abundance
ratio at later WC subtypes. This has persisted in the literature through to 
the present time  (e.g. Hirschi et al. 2005).
In contrast, stellar atmosphere models of WC4--8 stars have concluded that
there is no such trend with subtype (Koesterke \& Hamann 1995). Alternatively,
neglecting optical depth effects, recombination line theory may be applied
to hydrogenic transitions of C\,{\sc iv} in early WC stars (e.g. Smith \& Hummer 1998).
Since C\,{\sc ii-iii} transitions dominate the optical appearance of late WC 
stars, recombination line studies of late WC stars are extremely 
challenging.

The advent of sensitive mid-IR spectrometers aboard {\it Infrared Space
Observatory} and {\em Spitzer} allows alternative indicators of
evolutionary status, via fine-structure lines from elements sensitive to
advanced nucleosynthesis reactions, such as neon. Theoretically, at the
start of He-burning one expects the rapid conversion of $^{14}$N to
$^{22}$Ne, increasing the abundance of this isotope by a factor of
$\sim$100. Together with the cosmically more abundant $^{20}$Ne isotope,
one would expect a factor of $\sim$15 increase in the neon abundance.
Following techniques pioneered by Barlow et al. (1988), Ne/He in WC stars
may be determined from measuring fluxes of the [Ne\,{\sc ii}] 12.8$\mu$m
and [Ne\,{\sc iii}] 15.5$\mu$m lines, the latter solely observed from
space-based missions. Mid-IR studies with {\it ISO} have revealed high Ne
abundances in WC6--8 stars (Willis et al. 1997;  Dessart et al. 2000).  
The majority of late-type WC stars are known to form carbon-rich dust,
either persistent or episodically such that many are dust-dominated at
mid-IR wavelengths (Williams et al. 1987). Due to poor sensitivity of the
mid-IR {\it SWS} spectrometer, only dust dominated WC9 stars were observed
with {\it ISO}, completely veiling the stellar wind signature in the
mid-IR. To date, AS320 (WR121) is the only WC9 star for which a neon
abundance has been obtained, albeit based solely on [Ne\,{\sc ii}]
12.8$\mu$m (Smith \& Houck 2005).

In the current study, {\em Spitzer} IRS spectroscopy of HD~164270 (WR103,
WC9) is presented. HD~164270 does possess a dust shell (van der Hucht
2001), although this is sufficiently faint that stellar wind lines are
observed in the mid-IR, including the crucial fine-structure lines of Ne
and S. In order to derive neon and sulfur abundances, we perform the first
quantitative study of a WC9 star allowing for non-LTE radiative transfer
in an extended, clumped, expanding atmosphere allowing for metal line
blanketing (Hillier \& Miller 1998).

Finally, we carry out the first modern comparison between the chemical and
wind properties of HD~164270 and BD+30 3639 ([WC9], alias HD~184738,
Campbell 1893). The latter represents one of a small group of central
stars of Planetary Nebulae (CSPNe) which share the spectroscopic
appearance of WC stars.  Smith \& Aller (1971)  carried out an optical
qualitative spectroscopic comparison of these two stars and concluded that
they were virtually identical, except for narrower wind lines and strong
nebular lines in the CSPN.  BD+30 3639 is located in a young (800 yr)
compact nebula (4$''$ in diameter, Li et al. 2002)  whilst HD~164270, as
with most Population I WC stars, is not associated with an optically
visible nebula.   As with some dusty WC stars, PNe with late-[WC] 
type
central stars are very bright at mid-IR wavelengths due to hot dust within
their compact, dense nebulae, so were readily observed with {\it ISO/SWS}
(e.g. Waters et al 1998). In general, the material around massive
Wolf-Rayet stars known to have circumstellar nebulae may contain dust
condensed in one or more slow-moving outflows from previous mass-loss
episodes, or by interactions with the ISM.

Although both WC and [WC] stars involve hydrogen deficient stars at a late
stage of evolution, their evolutionary paths followed are remarkably
different.  WC stars represent the final phase of initially very massive
stars ($\geq 25M_{\odot}$ at Solar metallicity) undergoing advanced stages
of nucleosynthesis prior to core-collapse as a Supernova.  WC stars, with
ages of $\sim$5Myr, thus represent the bare cores of massive stars due to
extensive stellar winds and exhibit the products of core He-burning
(Meynet \& Maeder 2003).  In contrast, it is thought that [WC] stars
result from low initial mass ($\sim 2M_{\odot}$) progenitors following the
Asymptotic Giant Branch (AGB) phase. The hydrogen-rich stellar envelope of
AGB stars may be removed in a late thermal pulse following shell H and He
burning (Iben et al. 1983; Herwig et al. 1999), resulting in H-deficient
post-AGB stars, CSPNe and ultimately white dwarfs.

%__________________________________________________________________

\section{Observations}\label{obs}

Our study uses far-UV to mid-IR spectroscopy of HD~164270 and BD+30 3639,
of which {\em Spitzer} IRS mid-IR spectroscopy of HD~164270 is newly presented here.

\subsection{Mid-IR/near-IR spectroscopy}

HD~164270 was observed on 2004 March 4 (AOR key 6012160), 
with the {\em{Spitzer}}-IRS  (Houck et al. 2004) as part of the Guaranteed
Time Program WRDUST, using the long-slit Short-Low (SL) module at a spectral resolution
$R \equiv \lambda/\Delta\lambda \simeq 75-125$, and the Short-High (SH) and Long-High 
(LH) echelle modules each at 
$R \simeq 1000$. These cover the wavelength ranges
of $5.3 - 14.5~\mu$m (SL), 10--19.5~$\mu$m (SH), and 19--37.5~$\mu$m (LH). Once the
telescope's pointing was re-initalized to an accuracy of $\leq 1''.0$ (1~$\sigma$ 
radial) using
    high accuracy peakup on a nearby offset star with the 16 $\mu$m camera in the IRS, 
the 
    spectral observations were carried out in Staring Mode, which points the telescope 
    so that the star is observed at two separate positions along the length of each 
slit, in 
    principle allowing one to make background (zodiacal light and cirrus) emission 
corrections and 
    mitigating the residual effects of incident cosmic rays and solar particles.  The 
spectra were 
    obtained using total integration times of 84, 480, and 360 seconds in the SL, SH, 
and LH modules, 
    respectively.  Background and pixel corrections in the nodded observations works in 
practice 
    only for the SL observations, but cannot be achieved with the relatively short SH 
and LH 
    slits without dedicated off-source observations, which we did not acquire.  
Background 
    corrections to the SH and LH portions were therefore applied using the background 
model 
    available from the {\em{Spitzer}} Science 
    Center{\footnote{http://ssc.spitzer.caltech.edu/documents/background/}}, normalized 
to the 
    background levels observed with the SL portion at 10$\mu$m.  Corrections to outlier 
pixels
    are discussed below.

   Raw data cubes were processed on a pixel basis in the SSC pipeline (S10.5), which 
reformats 
    the pixel data and housekeeping information, removes dark currents and multiplexer 
drift, 
    applies corrections for inter-pixel photocurrent coupling and amplifier 
non-linearities, fits 
    the integration ramps and reduces the data cubes to 2-dimensional signal planes in 
units of 
    electrons/sec/pixel.  At this point we removed the background levels in the SL data, 
defined 
    approximately 1$'$ on either side of the star in the off-source subslits. The 
resulting illumination profile at 
    each wavelength was found to be consistent with that of a point source (no extended 
emission 
    was detected), and subsequent photometric point-source calibrations were applied 
during 
    spectral extraction.  Prior to extraction of the SH and LH data, however, we made 
systematic
    outlier pixel corrections using preceding dark current measurements obtained on the 
``blank'' 
    sky at the North Ecliptic Pole. The default super-darks subtracted in this version 
of the 
    pipeline are combined from measurements over several observing campaigns, but the 
    offline use of nearest-in-time dark currents provides a more accurate, 
contemporaneous match 
    to individual pixels with dark currents varying over several hours to a day.  
Remaining
    dead pixels in the illuminated regions of the arrays (totaling less than 5\%) were 
corrected
    by a spline interpolation scheme.  Spectra from each individual integration were 
then 
    extracted using offline SSC software, applying wavelength calibrations accurate to 
    $\sim$1/5 resolution element for well-pointed observations ($\pm$ 0.02 and 0.002 
$\mu$m in 
    SL and SH spectra at 10$\mu$m, and $\pm$ 0.005 $\mu$m in LH at 25 $\mu$m).  
Individual spectra 
    were then examined and combined.  Photometric uncertainties are estimated to be 5\% 
for SL, 
    5 -- 15\% for SH, and 15 -- 20\% for LH (increasing with wavelength due to the 
approximations 
    of the background levels). However, we find excellent agreement between our 
flux-calibrated
    IRS spectra and ground-based L, M and N-band photometry, discussed below.
    
% fig1

The IRS spectrum of HD~164270 is described by a blue stellar continuum,
with a number of spectral features superimposed.
We present selected regions of the IRS spectrum in Fig.~\ref{irs}, 
revealing a combination of spectral lines from He\,{\sc i} and C\,{\sc ii} 
formed in the inner stellar wind,  plus fine-structure lines from 
[Ne\,{\sc ii}] 12.8$\mu$m 
and  [S\,{\sc iii}] 18.7$\mu$m formed in the outer stellar wind,
with [Ne\,{\sc iii}] 15.5$\mu$m and [S\,{\sc iv}] 10.5$\mu$m notably
absent. We have measured the line fluxes from the fine-structure lines
to be 
$F_{\lambda}({\rm [Ne\,II]})$ = 2.18 $\pm 0.08 \times 10^{-12}$ erg/s/cm$^{2}$, and 
$F_{\lambda}({\rm [S\,III]})$ = 1.03 $\pm 0.01 \times 10^{-13}$ erg/s/cm$^{2}$.

For BD+30 3639 we use the combined ISO SWS/LWS spectrum from
Cohen et al. (2002), namely TDT35501531 (SWS01) and TDT35501412 (LWS01)
observed in Nov 1996,  to which the reader is referred for detailed 
information. The mid-IR continuum is reproduced by two temperature
blackbodies (290 and 98K) plus crystalline and amorphous silicate features
(Waters et al. 1998).
No stellar spectral features are observed due to the dominant dust signature.
[Ne\,{\sc ii}] 12.8$\mu$m  and  [S\,{\sc iii}] 18.7$\mu$m 
are also observed in BD+30 3639, although these are nebular in origin within
this system. The neutral nebular mass amounts to 0.13$M_{\odot}$ for
a distance of 1.2\,kpc as derived from the ISO/LWS [C\,{\sc ii}] 158$\mu$m
line flux (Liu et al. 2001).

We have also used near-IR spectroscopy of HD~164270 obtained with the 3.8m
UK Infrared Telescope (UKIRT) from August 1994. The cooled grating
spectrograph CGS4, 300mm camera, 75l/mm grating and a 62$\times$58 InSb
array were used to observe three non-overlapping settings, covering
1.03-1.13$\mu$m ($\lambda/\Delta \lambda$ = 800), 1.61-1.82$\mu$m ($\lambda/
\Delta \lambda$ = 600) and 2.01--2.21$\mu$m ($\lambda /\Delta \lambda$ = 800.
Details of the data reduction are provided by Crowther \& Smith (1996).
Eenens \& Williams (1994) derive a wind velocity of 1100 km/s using
their higher spectral resolution He\,{\sc i}  2.058$\mu$m observations of HD~164270.

% fig2

\subsection{UV/optical spectroscopy}

Flux calibrated optical spectroscopy of HD~164270 were obtained with
the 3.9m Anglo Australian Telescope and RGO spectrograph in November 1992
with a 1024x1024 Tektronix CCD, 1200V grating and 250mm camera, providing complete
coverage between $\lambda\lambda$3680--6000 in three overlapping settings at
1.7\AA\ resolution.
For BD+30 3639, flux calibrated spectroscopy was obtained at 
the 4.2m William Herschel Telescope, together with the dual-beam ISIS
spectrograph in August 2002. ISIS was equipped with an EEV CCD and 300B grating
(blue) and a Marconi CDD and 300R grating (red) plus the $\lambda$6100 dichroic, providing complete
spectral coverage between $\lambda\lambda$3400--8800 at 3.5\AA\ 
resolution.
% the 2.5m Isaac
% Newton Telescope together with the IDS spectrograph, Tektronix CCD, 1200B grating
% and  235 mm camera was used to provide complete coverage between $\lambda\lambda$3700--6800
% in four overlapping settings (1.5\AA\ resolution). 
A standard reduction was applied in both cases using
calibration arc lamps and spectrophotometric standard stars (e.g. see 
Crowther et al. 1998 for further details). Modern optical spectroscopy supports the findings
of Smith \& Aller (1971), with selected (classification) spectral regions presented in Fig.~\ref{compare}.

% fig3

UV high dispersion, large aperture
 spectroscopy of both targets was obtained from the $IUE$ Newly Extracted Spectra
(INES) archive, namely SWP8156 and LWP13972 for HD~164270, with SWP7594, 7642,
8591, 8863, 13333, 51870 and LWR6924, 7334 for HD~164270. Stellar wind velocities
($v_{\rm black}$) have been measured for Si\,{\sc iv} $\lambda$1394, revealing
1140$\pm$50 km/s for HD~164270 (Prinja et al. 1990) and 700$\pm$50 km/s 
for BD+30 3639. These values are adopted for the remainder of this analysis.

% 1140 km/s -> SiIV, 1090 km/s -> CIV 1550

% fig4

Finally, we also used {\it Far-Ultraviolet Spectroscopic Explorer (FUSE)} spectroscopy
of HD~164270 covering $\lambda\lambda$912--1190\AA, namely $FUSE$ dataset
P1171001 with an exposure time of 5.1ks, as discussed in Willis et al. (2004). 
Our adopted wind velocity of HD~164270 is reasonable agreement with the
average velocity of 1036 km/s in Willis et al. 

% \section{Analysis}\label{analysis}

\subsection{Distances and interstellar reddening}

We adopt spectrophotometry of HD~164270 from Torres-Dodgen \& Massey (1988), revealing
a narrow-band (Smith 1968) visual magnitude of $v$=8.86 mag, whilst we adopt $M_{v} = -4.6 \pm 0.4$ mag 
for WC9 subtypes, based on estimated absolute magnitudes for three Galactic WC9
stars from van der Hucht (2001). High resolution spectroscopy of several WC9 stars reveals
the characteristic signature of a luminous OB-type companion 
(Williams \& van der Hucht 2000), but no evidence of a luminous
companion to HD~164270 is known (Smith \& Aller 1971). From our spectral fits we
estimate $E(b-v)$=0.49  ($E(B-V)$=0.58) using R=$A_{\rm V}/E(B-V)$=3.0
which indicates a distance of $\sim$1.9\,kpc to
HD~164270. 

For BD+30 3639, our WHT spectrophotometry indicates $v=$10.28 mag.
Li et al. (2002) have recently analysed WFPC2 images obtained 5 yr 
apart to measure the 
PN expansion rate and so derive a distance of 1.2\,kpc which 
together with our estimate of $E(b-v)$=0.32
($E(B-V)=$0.39) indicates $M_{v} = -1.5$ mag. 

\subsection{Dust modeling}

De-reddened spectrophotometric flux distributions from 0.1 to 100$\mu$m
are presented in Fig.~\ref{wc9_lfl}, together with ground-based photometry from 2MASS, 
Williams et al. (1987) and Jameson et al. (1974), and
theoretical continua obtained from our analysis. 
These illustrate the UV/optical similarities, and mid-IR differences
between HD~164270 and BD+30 3639. 
Dust contributes negligibly to the total energy budget of HD~164270, whilst
the (re-radiated) dust in BD+30 3639 is comparable to the stellar flux.
We adopt a standard Galactic extinction law (R=3.1) for the 
UV/optical/near-IR (Howarth
1983), supplemented with a mid-IR extinction curve provided by M. Cohen 
(see Fig.2 of Morris et al. 2000).

Since the primary goal of the current study is to investigate the 
physical
and wind properties of WC9 stars, we use dust radiative transfer models 
from the recent literature. For HD~164270, we use a dust spectral energy 
distribution from TORUS (Harries, priv comm.) following the approach of 
Harries et al.  (2004) for WR104. The TORUS model assumes a spherical
shell composed of amorphous  carbon with a uniform size of 0.1$\mu$m and
a dust-to-gas ratio of 2$\times 10^{-5}$ by mass.  With respect to other 
very dusty WC9 stars (van der Hucht et al. 1996), the dust shell 
associated with 
HD~164270 is relatively weak, contributing at most 80\% of the continuum
flux at 3$\mu$m. 

For the dust associated with BD+30 3639 we use the MODUST 
model presented by Waters et al. (1998), who identified distinct (cooler) 
O-rich and (warmer)  C-rich shells. The MODUST model is successful in 
reproducing the observed energy distribution longward of 10$\mu$, where
dust contributes greater than 99.9\% of the continuum flux, but fails to
reproduce the 1--10$\mu$m excess. For both  HD~164270 and BD+30 3639, 
dust 
emission contaminates the stellar flux distribution longward of 1.3$\mu$m.

% fig5

\section{Quantitative analysis}

\subsection{Technique}

For the present study we employ CMFGEN (Hillier \& Miller 1998), which solves the 
transfer equation in  the co-moving frame subject to statistical and 
radiative equilibrium,   assuming an expanding, spherically-symmetric, homogeneous and static 
atmosphere, allowing for line blanketing and clumping. The stellar radius ($R_{\ast}$) is defined
as the inner boundary of the model atmosphere and is located at 
Rosseland optical depth of $\sim$20 with the stellar temperature ($T_{\ast}$)
defined by the usual Stefan-Boltzmann relation. 

Our approach follows previous studies (e.g. Crowther et al. 2002), such that
diagnostic optical lines of C\,{\sc ii} $\lambda$4267, C\,{\sc iii} $\lambda$5696
and C\,{\sc iv} $\lambda\lambda$5801-12
plus the local continuum level allow a determination of the stellar temperature, 
mass-loss rate and luminosity.  In contrast with earlier WC subtypes, a reliable
treatment of low temperature de-electronic recombination  C\,{\sc ii} lines is
necessary in WC9 stars. These lines are not included in Opacity Project calculations,
so improved {\sc superstructure} calculations were carried out by P.J.~Storey 
on our behalf. 

% fig6

Our final model atom contains He\,{\sc i-ii}, C\,{\sc ii-iv}, O\,{\sc ii-iv}, 
Ne\,{\sc ii-iv}, Si\,{\sc iii-iv}, S\,{\sc iii-vi}, 
Ar\,{\sc iii-v}, Ca\,{\sc ii-vi} and Fe\,{\sc iii-vi}. In total, 987 super levels, 3497
full levels and 40,952 non-LTE transitions are simultaneously considered.
We assume hydrogen is absent, with variable C, O and Ne abundances, and
Si--Fe initially fixed at Solar values (Grevesse \& Sauval 1998). As for
other WC subtypes, C/He abundances are ideally
obtained from He\,{\sc ii} $\lambda$5411 and C\,{\sc iv} $\lambda$5471.
Since these lines are weak in WC9 stars, the accuracy with which C/He
may be obtained in this way is reduced relative to earlier subtypes, but
should be accurate to $\pm$0.1 by number. 
For oxygen, O\,{\sc ii} 
$\lambda$4415--4417, O\,{\sc iii} $\lambda$3754--3759, $\lambda$5592
provide the cleanest oxygen diagnostics in the visual region, and should
permit a determination of O/He to within a factor of two. Williams
\& van der Hucht (2000) revealed that WC9 stars span a range in O\,{\sc 
ii} $\lambda$4415-4417 line strength, suggesting a factor of $\geq$2 
spread in oxygen abundance. 

We adopt a standard form of the velocity law,
$v(r) = v_\infty ( 1 - R_*/r)^\beta$, where $\beta$=1. 
The mass-loss rate is actually derived as the ratio $\dot{M}/\sqrt{f}$,
where $f$ is the volume filling factor that can be constrained by
fits to the electron scattering wings of the helium line profiles 
(following Hillier 1991)  resulting in $f\sim$0.1. From the observed strength
of electron scattering wings we can definitely 
exclude homogeneous mass-loss in both HD~164270 and BD+30 3639. 
% In addition, the mid-IR continuum slope also reacts to different filling-factors.

\subsection{Physical and Wind properties}\label{spect}

We compare our synthetic model of HD~164270 to  de-reddened
UV ($IUE$) and optical (AAT) spectroscopic observations in Fig.~\ref{wr103_sp}. 
The agreement between the spectral features and continuum is generally excellent. 
The majority of synthetic 
near-UV and  optical (primarily He, C, O) lines match observations to 
within  $\sim$30\%. The only exception is C\,{\sc iii} $\lambda$4647--50
which is too weak. This can be resolved with a 5\% higher 
temperature, at the expense of other fits to optical lines.
%
% Exceptions include C\,{\sc ii} $\lambda$2325, 
%C\,{\sc  iv}  $\lambda$2405, the complex around C\,{\sc iv} 
%$\lambda$2530 and C\,{\sc iii} $\lambda$4650 which are each predicted
%too  weak in emission. 
%
In the far-UV, numerous Fe features  
dominate the spectral  appearance, together with C and Si resonance lines. 
The synthetic Fe spectrum is generally too strong is emission, as are
Si\,{\sc iii} $\lambda$1295--1312 and
the blue component of Si\,{\sc iv} 
$\lambda$1393--1402, whilst C\,{\sc iii} $\lambda$2297  is 30\% too weak. 
C\,{\sc iv} $\lambda$1548--51 is too weak in emission at a stellar 
temperature of 48kK.  As with C\,{\sc iii} $\lambda$4647--51, this can be 
reproduced at a slightly higher temperature, albeit with poorer agreement
in the iron forest around $\lambda$1450. 

We compare the de-reddened 
$FUSE$  far-UV  spectrum of HD~164270 to the synthetic spectrum in 
Fig.~\ref{wr103_fuse}. Agreement is reasonable longward of $\lambda$1120, 
where few molecular H$_2$
lines absent, including the Si\,{\sc iv} $\lambda$1122--28 doublet, whilst
Lyman and Werner molecular bands dominate the spectrum at shorter wavelengths, as 
demonstrated by the apparent absence of the C\,{\sc ii} $\lambda$1037 
in the $FUSE$ dataset (see also Willis et al. 2004).

% Near-IR spectroscopy of HD~164270 is compared to
% the synthetic spectrum  in Fig.~\ref{wr103_ir}, allowing for hot dust 
% (Harries, priv. comm.).
% At 1$\mu$m there is negligible contribution by hot dust, with He\,{\sc 
% i} 1.0830$\mu$m
% well reproduced. At 2$\mu$m, the dust contribution is significant, with 
% a reasonable
% comparison between observations and the synthetic spectrum at He\,{\sc i} 2.0581$\mu$m. 
% Dust contributes
% the majority of the continuum in the mid-IR (recall Fig.~\ref{wc9_lfl}) with spectral
% features of He\,{\sc i} and C\,{\sc ii} typically too strong, although the disagreement may
% be due to the fractional contribution of dust continuum rather than the stellar model.

For completeness, we compare UV ($IUE$) and optical (WHT) 
spectrophotometry
of BD+30 3639 to our synthetic model in Fig.~\ref{bdp30_sp}. Comments are as
for HD~164270, except for the strong nebular emission superimposed upon the
stellar spectrum.
The stellar parameters derived for each star are presented in Table~\ref{table1}.
The derived stellar temperature of BD+30 3639  is higher than HD~164270
due to its smaller physical size, whilst their wind efficiencies, 
 $\eta = \dot{M} v_{\infty}/ (L/c)$ are effectively identical, despite 
stellar masses a factor of ten different. Mass estimates originate from
the H-free mass-luminosity relation of Schaerer \& Maeder (1992) for
HD~164270, and Herwig et al. (1999) for BD+30 3639.

Recent  theoretical (Herwig et al.  2003) and observational 
(Reiff et al. 2005) evidence suggests that  H-deficient PG1159 
stars (for which [WC] subtypes  are likely precursors) are severely Fe-depleted.
Thus far we have adopted Solar (Grevesse \& Sauval 1998) heavy metal abundances.
Consequently, we additionally considered models in which the iron content is reduced by a 
factor of ten with respect to Solar. 
This reduced iron model compares rather more favorably to observation
in the  $\lambda$1400--1700 region (Fe\,{\sc iv-v}), whilst the Solar Fe 
model is in better agreement for $\lambda \geq$1800 (also Fe\,{\sc iv}). 
Recall, however that the synthetic far-UV spectrum for HD~164270 was
also rather too strong with respect to observations, so a more robust
result awaits improved spectroscopic fits across the entire far-UV.
Fortunately, the  question of Fe depletion in BD+30 3639 does
not affect our main aim of the present study, since the optical 
spectroscopic appearance of the Solar and Fe-depleted models are virtually identical.

\subsection{Comparison with previous results}

% fig7

Our derived stellar temperature of $T_{\ast} \sim$ 48kK for HD~164270 
is rather higher than
the previous estimate of 30kK for WC9 stars, based on pure helium model atmospheres 
(Howarth \& Schmutz 1992). This follows the general trend in which the inclusion
of line blanketing leads to an increase in stellar temperatures for WR stars 
(Crowther 2003b).  The present study represents the first detailed investigation 
of any WC9 star, it is in good agreement with the  estimate of 45kK for WR104 
by Crowther (1997) using a He, C and O  non-LTE model. 

Since late-WC stars are preferentially dust formers, we note that the ionizing
flux distribution of HD~164270 is rather harder than those commonly adopted in
chemical and radiative transfer studies of dust in such environments (e.g. Williams et 
al. 1987; Cherchneff et al. 2000). One notable exception is the 3D dust radiative transfer study of 
WR104 (Pinwheel nebula) by Harries et al. (2004) for which a similar stellar
spectral energy distribution to the current study was adopted. The origins of dust around 
Population~I late WC stars is still under debate, although recent studies suggest a  binary system 
is required to provide H-rich  material and the necessary shielding from the  intense radiation for 
grain formation (Crowther 2003a). To date, there is no evidence for a luminous companion to 
HD~164270 (Smith \& Aller 1971).

BD+30 3639 has been studied by Leuenhagen et al. (1996) using a non-LTE
model atmosphere study, also allowing for He, C and O, such that blanketing
and clumping were neglected. As with HD~164270,  we obtain 55kK 
versus
47kK from Leuenhagen et al. (1996). However, blanketing tends to redistribute high energy photons to
lower energy, leading to a softer ionizing flux distribution. In Table~\ref{table1} we
include the Lyman and He\,{\sc i} continuum ionizing fluxes for both stars. The relatively 
soft radiation explains why previous nebular techniques have obtained substantially lower
estimates of the stellar temperature (e.g. 34kK: Siebenmorgen et al. 1994). 
In contrast to the 
amorphous carbon nature of dust associated with late WC stars (Williams et al. 1987),
 the PN associated with BD+30 3639 exhibits crystalline silicate dust, signaling
a recent change from an oxygen- to carbon-dominated chemistry (Waters et al. 1998). 
De Marco et al. (2003) argue binarity could naturally explain the dual dust
chemistry of PN and the H-depletion of the CSPN. 

Finally, if we scale the Leuenhagen et al result for BD+30 3639 to our
assumed distance of 1.2\,kpc, we obtain a revised luminosity of $\log
L/L_{\odot} = 4.0$, versus 3.8 obtained here, and a (homogeneous)  
mass-loss rate of 4$\times 10^{-6} M_{\odot}$ yr$^{-1}$ versus our
estimate of 3$\times 10^{-6} M_{\odot}$ yr$^{-1}$.

\subsection{Higher radio derived mass-loss rates for WC9 stars}\label{radio}

Leitherer et al. (1997) obtained an  upper limit to the
6cm radio flux for HD~164270 of
$<$0.42 mJy, equating to a homogeneous
mass-loss rate
of $< 2.7 \times 10^{-5} M_{\odot}$ yr$^{-1}$  
under the {\it assumption} of singly ionized He, C and O for the
radio forming continuum, versus $\dot{M}/\sqrt{f} \sim
3 \times 10^{-5}  M_{\odot}$  yr$^{-1}$  obtained here
from an analysis of the UV, optical and near-IR spectrum. 
Since the radio estimate is recognized as
providing the least model dependent mass-loss rates of 
Wolf-Rayet stars, does the present study
contradict the radio results?

Fig.~\ref{ion} presents the theoretical wind structure of our 
HD~164270 model,
including the variation of electron temperature, electron density, velocity and 
stellar radius versus Rosseland optical
depth, together with the ionization balance of the principal elements, namely 
He, C, O, Fe. We also include Ne and S 
since we consider their ionization state in the outer winds in Sect.~\ref{neon}. 
The winds of WC9 stars are highly stratified, such that for example 
S$^{6+}$ is the dominant
ion of sulfur in the inner wind ($\sim$10$^{14}$ cm$^{-3}$), 
whilst S$^{2+}$ is the dominant ion in the outer wind ($\sim$10$^{7}$ cm$^{-3}$).

From the figure, our model atmosphere supports the dominant ionization stages
of C$^{+}$ and O$^{+}$ in the outer wind, whilst He is partially recombined 
beyond several hundred $R{\ast}$. Therefore, the claimed lower limit to the radio 
mass-loss rate by Leitherer et al. (1997) will increase.  In fact, our 
model atmosphere predicts 2cm and 6cm radio fluxes of only 0.2mJy and 0.09mJy, respectively,
for HD~164270. Similar  arguments  apply to the  apparently low mass-loss rates of WC9 
stars derived from radio  observations by Cappa et al. (2004). As a 
consequence, their
conclusions  that WC9 stars possess weaker winds than earlier subtypes is not necessarily 
valid.

% table 1 in text

\subsection{Elemental abundances}\label{neon}

We admit a greater uncertainty in C/He for WC9 stars
with respect to earlier WC stars due to the weakness of 
the diagnostic He\,{\sc ii} $\lambda$5411
and C\,{\sc iv} $\lambda$5471 lines. We obtain C/He$\sim$0.2$\pm$0.1
for HD~164270 and C/He$\sim$0.25$\pm$0.1 for BD+30 3639, by number.
O\,{\sc ii} lines at $\lambda$4415--4417, O\,{\sc iii} 
$\lambda$3754--3759, $\lambda$5592
represent the primary abundance diagnostics, revealing O/He$\sim$0.01
for HD~164270 and O/He$\sim$0.05 for BD+30 3639, which ought to be
reliable to a factor of two. 
No abundance studies have previously been reported for HD~164270,
whilst Leuenhagen et al. (1996) derived C/He=0.37 and O/He=0.03, by number
for BD+30 3639.

The fine-structure lines of [Ne\,{\sc ii}] 12.8$\mu$m
and [S\,{\sc iii}] $\lambda$18.7$\mu$m observed in HD~164270 (Fig.~\ref{irs})
permit the determination of ion fractions of Ne$^{+}$ and S$^{2+}$
in the outer wind of HD~164270. The critical densities for these fine-structure
lines ($\sim10^{5}$ cm$^{-3}$) relate to the constant velocity regime at
several
thousand stellar radii in WC9 stars, comparable to the radio forming continuum.
We determine abundances using numerical techniques introduced by Barlow et al. (1988),
adapted to account for a clumped wind by Dessart et al. (2000) and Smith \& Houck (2005).
With regard to clumping, the volume filling factor within this region may be quite distinct
from that estimated using optical recombination lines (Runacres \& Owocki 2002).

The numerical expression for the ion number 
fraction, $\gamma_i$, in a clumped medium (with volume filling
factor $f$) is (in cgs units)
\begin{equation}
 \gamma_{i}=\frac{(4 \pi \mu m_{H} v_{\infty})^{1.5}}{\ln(10)f^{0.25}}
\left(\frac{\sqrt{f}}{\dot{M}}\right)^{1.5}
\frac{1}{F_{u}(T)}\frac{2D^{2}I_{ul}}{\sqrt{\gamma_{e}}A_{ul}h\nu_{ul}} 
\end{equation}
where  $D$ is the stellar distance, $I_{ul}$ is the line flux of the 
transition with energy $h\nu_{ul}$ between upper level $u$ and lower 
level $l$, with transition probability $A_{ul}$.
$\gamma_e$ (=1.0) and $T_e$ (=6000K) are the electron number density
and temperature in the line-forming region, and the mean molecular weight
is $\mu$ (=5.5) for HD~164270, with $m_H$ the mass of the hydrogen atom. The
integral part, $F_u(T)$, is
\begin{equation}
F_{u}(T) = \int_{0}^{\infty}\frac{f_{u}(N_{e},T)}{\sqrt{N_{e}}}d 
\log(N_{e}).
\end{equation}
where $f_u$ is the fractional population of the upper level.

% fig8

% analytical 2.1e-3 Ne+/He and 5.4e-5 S2+/He

Following this technique, we derive $\gamma_{Ne^{+}} = 1.8\times 10^{-3}$
and $\gamma_{S^{2+}} = 4.2\times 10^{-5}$. Consequently,
Ne$^{+}$/He = 2.2$\times  10^{-3}$ and S$^{2+}$/He = 5.1$\times 10^{-5}$ by number. 
Of course, one needs to account for potential contributions from unseen
ionization stages. From Fig.~\ref{ion}, our model atmosphere predicts that
Ne$^{+}$ and S$^{2+}$ are the dominant ions of neon and sulfur at
the outer model radius of 1000$R_{\ast}$, where
$\log (n_{e}/$cm$^{3})\sim$6.7.
A similar degree of ionization at  $\log (n_{e}/$cm$^{3})\sim$4.5--5.8, where the 
[S\,{\sc iii}] and [Ne\,{\sc ii}]
fine-structure lines form, is supported by the absence of the
[S\,{\sc iv}] 10.5$\mu$m and [Ne\,{\sc iii}] 15.5$\mu$m fine-structure lines,
although we cannot exclude contributions from Ne$^{0}$ and S$^{+}$.

Consequently, we can set a lower limit to the Ne/He ratio with
Ne$^{+}$/He = 2.2$\times 10^{-3}$. This implies a minimum
mass fraction of neon of 0.7\%, a factor of 7 times {\it higher} than the Solar
value of Ne/H $\sim 7.2 \times 10^{-5}$ by number,
as recently re-derived by Asplund et al. (2004) and Lodders 
(2003)\footnote{Note that the Bahcall et al. (2005) propose a higher
Ne abundance of Ne/H=1.9$\times 10^{-4}$ by number in order to resolve
the downward revised oxygen content of the Sun by Asplund et al. (2004)  
with helioseismology measurements.}. For sulphur, we can equally set
a minimum abundance of sulfur with S$^{2+}$/He = 
5.1$\times$10$^{-5}$. This indicates a minimum sulfur 
mass fraction of 0.024\%, a factor of two times {\it lower} than the 
Solar mass fraction, with an unknown fraction singly ionized.

Smith \& Houck (2005) have recently estimated the Ne abundance of AS320 (WR121,
WC9d) based on ground-based [Ne\,{\sc ii}] 12.8$\mu$m spectroscopy. They
estimated Ne$^{+}$/He = 4.1$\times 10^{-3}$, based on the analytical approach
of Barlow et al. (1998), a mass-loss rate 
that was 0.1~dex lower than a radio derived upper limit (though recall  Sect~\ref{radio}),
plus average C/He abundances for WC8--9 stars from the literature. The more accurate
numerical technique applied above typically leads to a 10--20\% reduction in
ion abundance relative to the analytical approach.  Since [Ne\,{\sc iii}]
15.5$\mu$m is unavailable from the ground, Smith \& Houck (2005) inspected
our HD~164270 {\em Spitzer} IRS dataset, which confirmed that Ne$^{+}$ instead
of Ne$^{2+}$ is the dominant ion in the outer wind of WC9 stars.

%\subsection{Iron depletion in BD+30 3639?}

\section{Discussion and Conclusions}

 We have presented the first quantitative, multi-wavelength 
spectroscopic study of
a massive WC9 star, HD~164270. Overall, the non-LTE model is remarkably 
successful at reproducing the far-UV to mid-IR spectrum of HD~164270
such that we can be confident in the resulting physical and chemical
properties.  

With respect to earlier subtypes, WC9 stars genuinely possess
lower stellar temperatures. However,  we find a similar 
C/He for HD~164270 to earlier WC subtypes, based on identical abundance diagnostics, 
contrary to the decreasing C/He ratio for later WC subtypes proposed by
Smith \& Maeder (1991). This is emphasised in Fig.~\ref{abundances} which compares
the carbon and neon abundances of all Galactic WC stars analysed by Dessart et al.
(2000) to which we have added HD~164270. Chemically, WC9 stars therefore appear normal with 
respect to earlier WC subtypes. This conclusion is supported by Smith \& 
Houck (2005) who
estimated  Ne/He=4.1$\times 10^{-3}$ for another WC9 star, AS320, using ground-based 
spectroscopy, albeit with an uncertain, radio-derived mass-loss rate.
Indeed, we shall
suggest below that the presence of a WC dense wind intrinsically leads to a late-type
WC star, so one might actually expect a substantial range of C/He within WC9 stars, as
suggested by Williams et al. (2005).

In Fig.~\ref{abundances} we also include the predicted neon versus 
carbon
enrichment during the WC stage for an initial 60$M_{\odot}$ star at
Z=0.020 initially rotating at 300 km/s from Meynet \& Maeder 
(2003). HD~164270 supports the previous results from Dessart et al.
(2000) that measured Ne enrichments fall a factor of 2--3 times lower
than predictions. During the early WC phase it is $^{22}$Ne 
enrichment by up to a factor of 200, rather than the initially more
abundant $^{20}$Ne which dominates the predicted enrichment. This
increase is dictated by reaction rates and  is unrelated to rotation 
(Meynet \& Meynet 2003). Considering this poor agreement, a 
{\em Spitzer} guest investigator program is presently underway to 
study a larger sample of WC and WO stars to assess whether the
relatively modest Ne enrichments obtained to date are typical.

Crowther et al. (2002) concluded that the defining characteristic 
distinguishing mid and early WC subtypes
 was primarily due to a higher wind density in the former,
rather than significantly lower stellar temperature. One would be able to
readily explain the observed preference of WC9 stars 
to inhabit high metallicity 
regions  (e.g. Hopewell et al. 2005; Hadfield et al. 2005) if the wind strengths 
of  WC stars increased with metallicity (Vink \& de Koter 2005), 
i.e.  high wind densities produce late subtypes, without the need for 
reduced stellar temperatures. 

In contrast, radio mass-loss determinations of WC9 stars by Leitherer et al. (1997) and
Cappa et al. (2004) suggest that they possess relatively low wind densities. 
However, helium is assumed to be singly ionized in radio mass-loss rate 
determinations, whilst Fig.~\ref{ion} indicates that helium is partially recombined  in 
the outer winds of WC9 stars. 
Consequently, previous radio mass-loss rate estimates for
WC9 stars represent lower (rather than upper) limits.

In Fig.~\ref{mdot} we present the mass-loss  rate versus luminosity
relationship obtained for Galactic Wolf-Rayet stars by Nugis \& Lamers
(2000) for an assumed metal content of C/He=0.25 and C/O=5 by number,
together with Galactic and LMC spectroscopic results from Crowther et al.
(2002) and references therein, plus HD~164270. The generic relation of
Nugis \& Lamers is very successful at reproducing the mass-loss rates of
Galactic WC5--7 stars. LMC WC4 stars possess low mass-loss rates for their
high luminosities and $\gamma$ Vel (WC8+O7.5\,III) and HD~164270 (WC9)  
possess high mass-loss rates for their low luminosities. Despite our small
sample, one might conclude that late WC stars result from high wind
densities instead of low stellar temperatures.  However, stellar
temperatures of $\gamma$ Vel (De Marco et al. 2000) and HD~164270 are the
lowest of the entire sample, so the low ionization of WC8--9 stars appears
to result from a combination of both high wind density and low stellar
temperature.

Note that there is also an apparent trend for higher luminosities (masses) at 
earlier subtype in Fig.~\ref{mdot},  although the reduced mass-loss rates 
prior to the Wolf-Rayet phase for the LMC stars relative to the 
Milky Way  counterparts probably dominates this difference. Nevertheless, 
Galactic WC5--7 stars reveal higher present masses than WC8--9 stars 
(recall all have well established distances with the exception of
HD~164270).

The recent suggestion of severe Fe-depletion in the likely precursors of late [WC] stars 
(Herwig et al. 2003; Reiff et al. 2005) potentially  impacts upon  the 
ability of the star 
to exhibit a strong wind. Indeed,  Gr\"{a}fener \& Hamann (2005) and Vink \& de Koter (2005)
both argue that winds of Wolf-Rayet stars are driven by radiation  pressure with Fe-peak elements 
the principal components of the line driving.  Consequently, a significant Fe depletion in 
BD+30 3639 might potentially reduce its wind  density to significantly below that of HD~164270. 
Without the line driving from Fe-peak elements, [WC] subtypes may either be unlikely to occur, or 
possess  very different wind properties from their Population~I cousins.
In contrast, we have demonstrated quantitatively that the winds of 
HD~164270 (WC9) and BD+30 3639 ([WC9]) are remarkably similar, supporting the earlier
descriptive result from Smith \& Aller (1971), despite very different evolutionary states,
arguing against a severely depleted Fe-content in BD+30 3639.

%In the single
%star scenario,  a late thermal pulse occured (at most) a few thousand years ago, shortly before BD+30 3639
%left the AGB (Herwig et al. 1999). 

%\subsection{}

% Stenholm - long period binary - disk? [Moffat question following Williams IAU 163 p.345)

% Analysis of Spitzer-IRS mid-IR fine structure lines of 
% Ne and S formed in the outer stellar wind of HD~164270 (in contrast to nebular origin for BD+30 3639)
% indicate significant Ne enrichment, with S potentially sub-Solar.

%Within the context of [WC] stars, our result together with other stars presented 
%in  Crowther et al. (2003) 
%resolves the previous inconsistency between early and late [WC]
%subtypes. Evolution was thought to proceed from late to early [WC] 
%with subsequently evolution to PG1159 stars and DO white dwarfs.  
%However, derived surface C+O abundances previously showed a {\it decrease} 
%from  late  to early (Koesterke \& Hamann 1997). Our results, allowing for 
%line  blanketing, indicate similar (C+O)/He abundances in late and early 
%[WC]  subtypes. Finally, the question of potentially reduced Fe in [WCL] 
%stars  is  considered. Some spectroscopic evidence argues in favour of a 
%reduced Fe content, yet if both [WC] and WC winds {\it are} radiatively 
%driven by  $\alpha$ {\it and} Fe peak elements, one might expect reduced 
%wind densities in the former if Fe is genuinely depleted, which is not
%the case for BD+30 3639 versus HD~164270.

Finally, we should emphasize that the physical and wind properties 
derived here for
HD~164270 and BD+30 3639 are not based on self-consistent
hydrodynamical models, i.e. mass-loss rates are imposed. Great progress 
towards such fully consistent models for Wolf-Rayet stars is now underway 
towards self-consistent wind models (notably Gr\"{a}fener \& Hamann 2005). 
In the meantime, from our comparison between WC and [WC] stars it appears 
that the combination of a hydrogen-deficient, metal-rich composition and
intense radiation field are sufficient conditions for the development of a
strong stellar wind, regardless of the evolutionary history (or present
mass) of the star.

\begin{acknowledgements}
Thanks to John Hillier for providing CMFGEN and making modifications to 
incorporate  new C\,{\sc ii} {\sc superstructure} calculations by Pete
Storey. Martin Cohen kindly provided the ISO spectrum of 
BD+30 3639. We also thank Tim Harries for calculating a dust model 
for 
HD~164270 with TORUS and Alex de Koter for providing the MODUST dust model
for BD+30 3639. PAC acknowledges financial support from the Royal 
Society.
This work is based in part on observations made with the WRDUST program (Principal
Investigator J. Houck), using {\em{Spitzer}} Space Telescope, 
which is operated by the Jet Propulsion Laboratory, California Institute of Technology 
under NASA contract 1407. Support for this work was provided by NASA through an award 
issued by JPL/Caltech. This study also makes use of data products from 
the Two Micron All Sky Survey, which is a joint project of the University of Massachusetts and 
IPAC/CalTech, funded by NASA and the NSF, and from the NASA/IPAC Infrared
Science Archive, which is operated by JPL/CalTech, under contract with NASA.
\end{acknowledgements}

\clearpage

\begin{table}[ht]
\caption[]{Stellar and wind properties of HD~164270 and BD+30 3639,
including Lyman ($Q_{0}$) and HeI ($Q_{1}$) continuum ionizing photons.
\label{table1}}
\begin{small}
\begin{tabular}{lrrrr}
\noalign{\hrule\smallskip}
Star & HD~164270 & BD+30 3639 \\
Sp Type     & WC9 & [WC9] \\
\noalign{\hrule\smallskip}
$T_{\ast}$ (kK)      & 48     & 55\\
$T_{\tau = 2/3}$ (kK)   & 41     & 48\\
$\log (L/L_{\odot})$ & 4.90   & 3.78 \\
$R_{\ast}/R_{\odot}$        & 4.1    & 0.85\\
$\dot{M}/\sqrt{f} (M_{\odot}$yr$^{-1}$) & --4.50 & --5.55\\
$f$                 & 0.1     & 0.1   \\
$v_{\infty}$ (km/s) & 1140 & 700 \\
$\dot{M} v_{\infty}/(L/c)$ & 7 & 5 \\
C/He                & 0.20     &  0.25   \\
O/He                & 0.01     &  0.05   \\
$M (M_{\odot})$     &   6      & 0.6: \\
$\log Q_{0}$ (s$^{-1}$)   & 48.57   & 47.50 \\
$\log Q_{1}$ (s$^{-1}$)   & 40.38   & 40.28 \\
$M_{v}$ (mag)       & --4.6   & --1.5 \\
\noalign{\hrule\smallskip}
\end{tabular}
\end{small}
\end{table}

\clearpage

\begin{figure}
\begin{center}
\includegraphics[width=1\columnwidth,clip,angle=0]{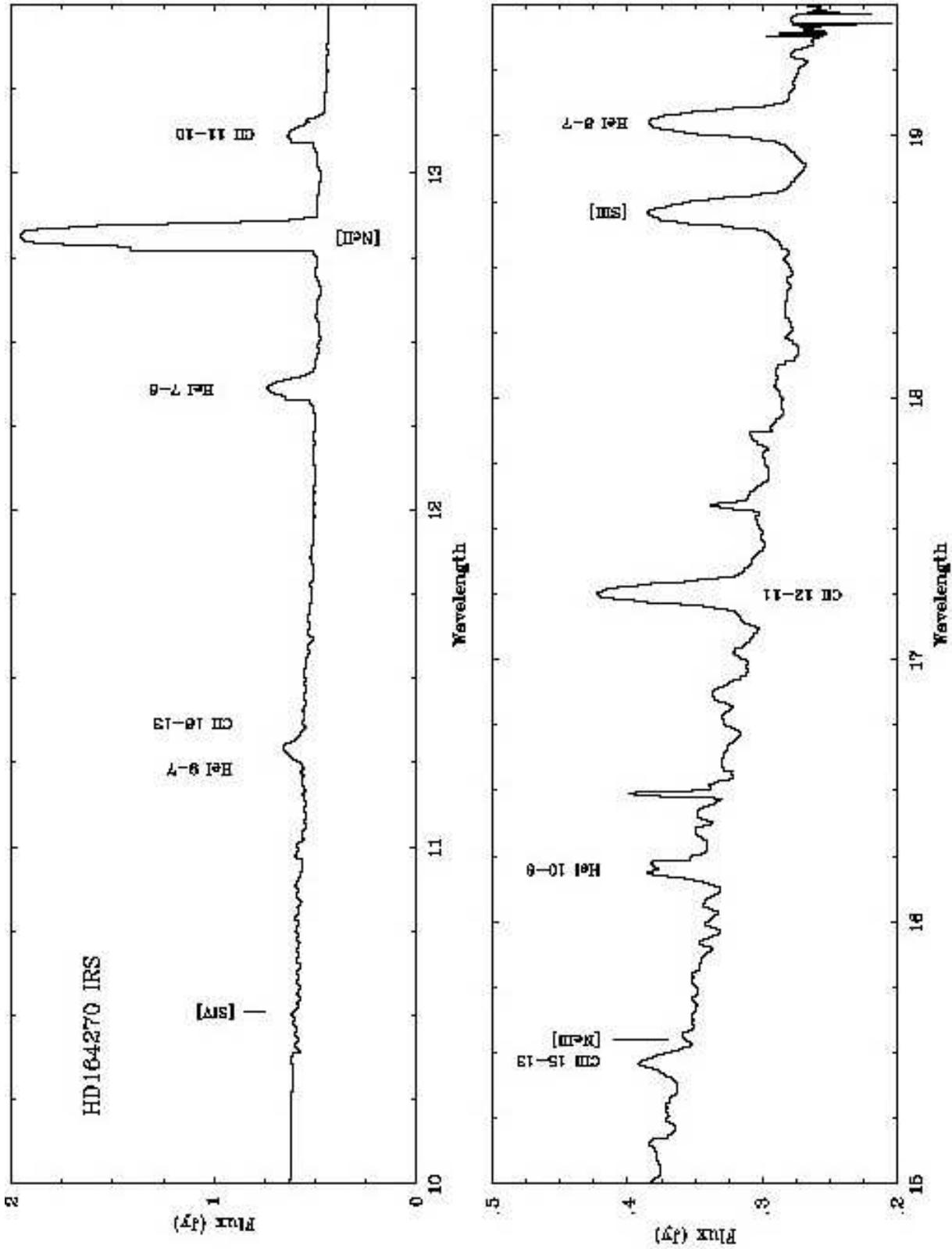}
\end{center}
\caption{Selected regions of the IRS  Short-High (SH) echelle spectrum 
(R$\simeq$1000) for HD~164270 (WC9).\label{irs}}
\end{figure}

\begin{figure}
\begin{center}
\includegraphics[width=1\columnwidth,clip,angle=0]{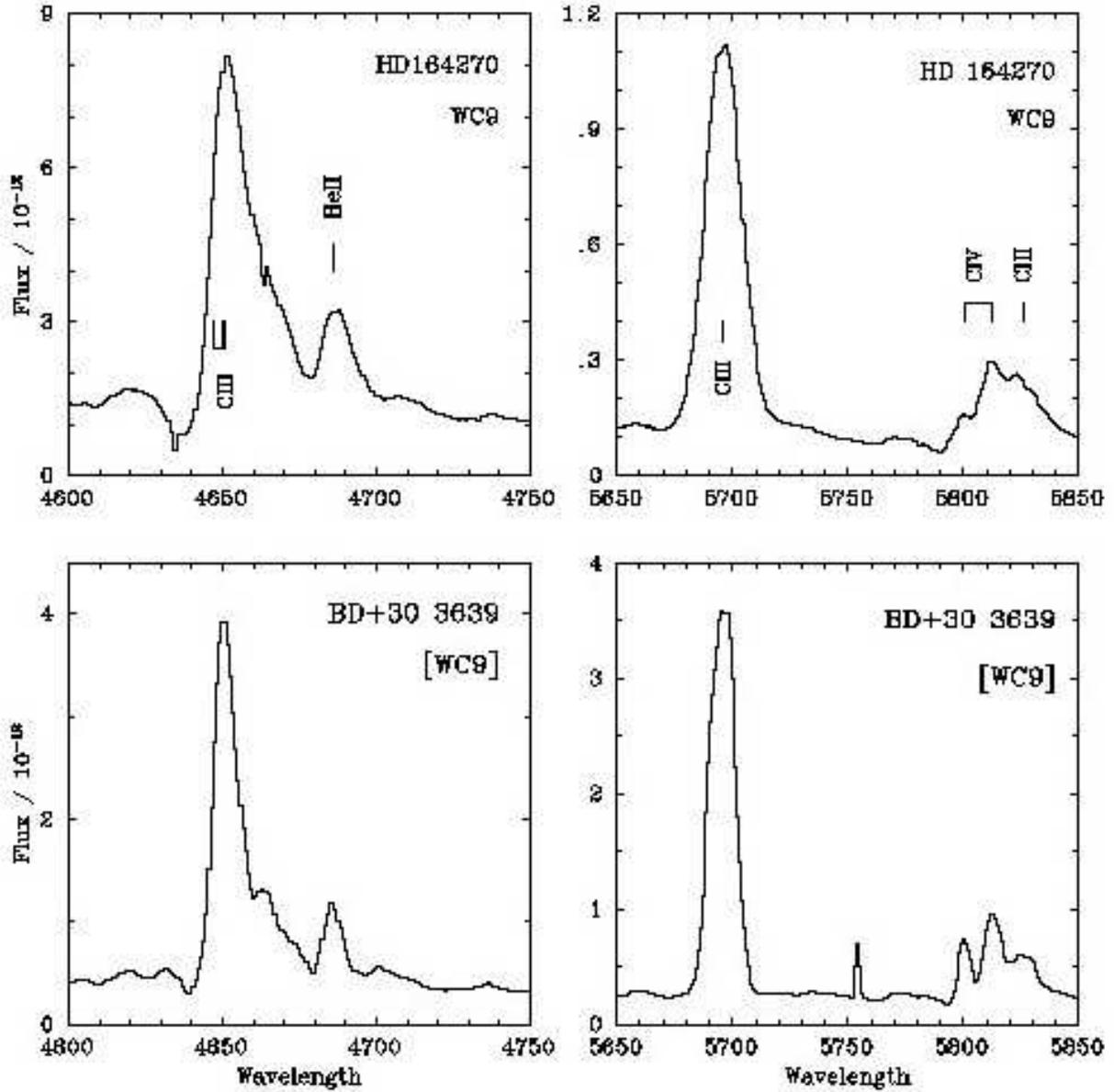}
\end{center}
\caption{Spectroscopic comparison between HD~164270 (AAT/RGO, 1.7\AA\
resolution) and BD+30 3639 (WHT/ISIS, 3.5\AA\ resolution) in
the vicinity of C\,{\sc iii} $\lambda$4650/He\,{\sc ii} $\lambda$4686 and
C\,{\sc iii} $\lambda$5696/C\,{\sc iv} $\lambda$5801--12, indicating a
very similar appearance except for narrower lines in BD+30 3639, 
originating from a slower stellar wind.\label{compare}}
\end{figure}

\begin{figure}
\begin{center}
\includegraphics[width=0.8\columnwidth,clip,angle=0]{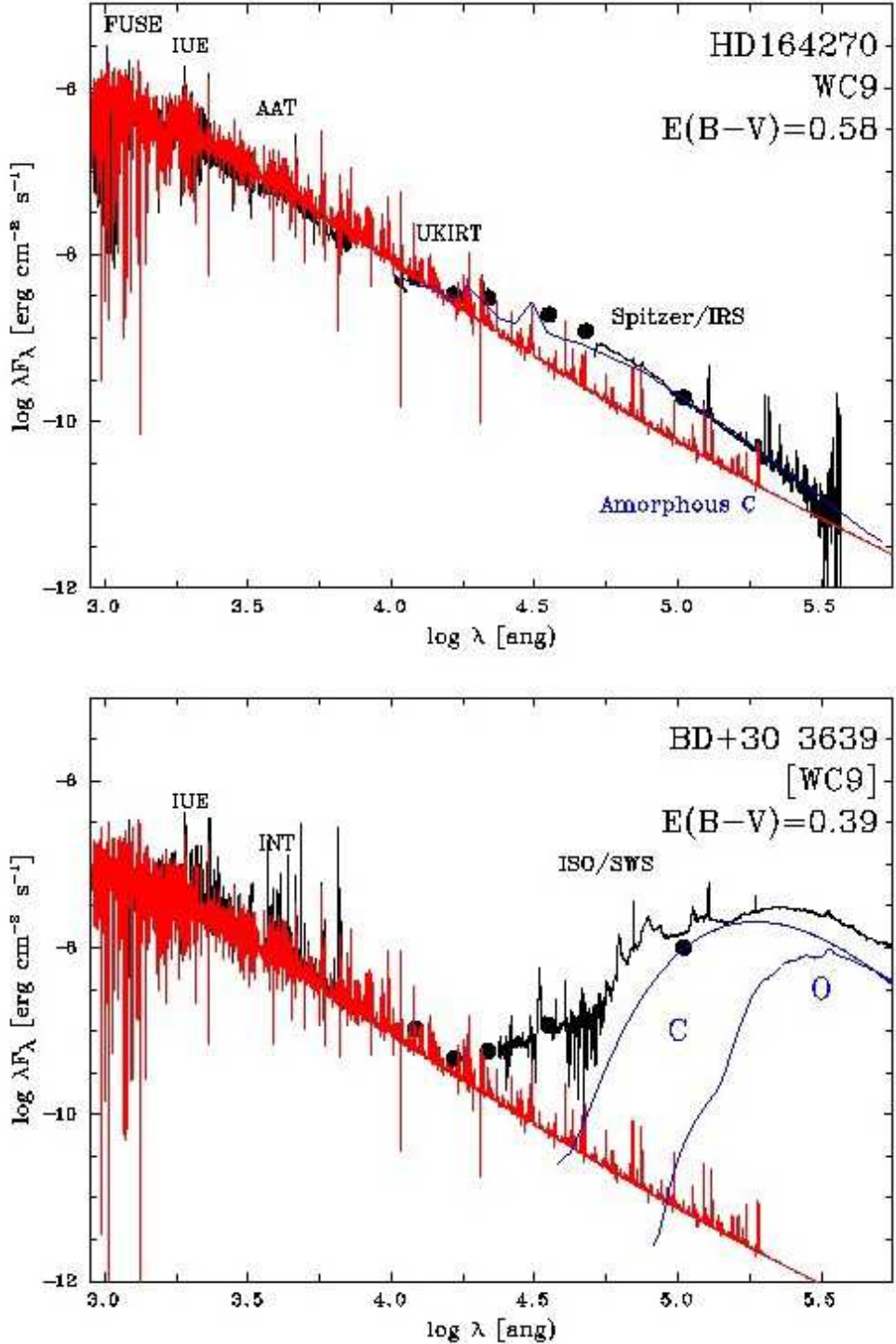}
\end{center}
\caption{De-reddened spectral energy distributions (erg\,s$^{-1}$\,cm$^{-2}$)
from 0.1-100$\mu$m for HD164270 (upper
panel) and BD+30 3639 (lower panel), including ground-based JHKLMN photometry (solid circles), 
together with stellar  continua from our analysis (red), dust  
distributions (green) from TORUS (Harries, priv. comm.) and MODUST 
(Waters 
et al. 1998). 
In both cases the stellar continuum dominates the  flux distribution in the
UV/optical, with dust contributing longward of 1$\mu$m. At 10$\mu$m the circumstellar dust in HD~164270 provides 
$\sim$70\% of the continuous flux, whilst the re-radiated dust from the PN 
of BD+30 3639 provides 99.9\% of the continuous flux.
\label{wc9_lfl}}
\end{figure}

\begin{figure*}
\begin{center}
\includegraphics[width=0.85\columnwidth,clip,angle=0]{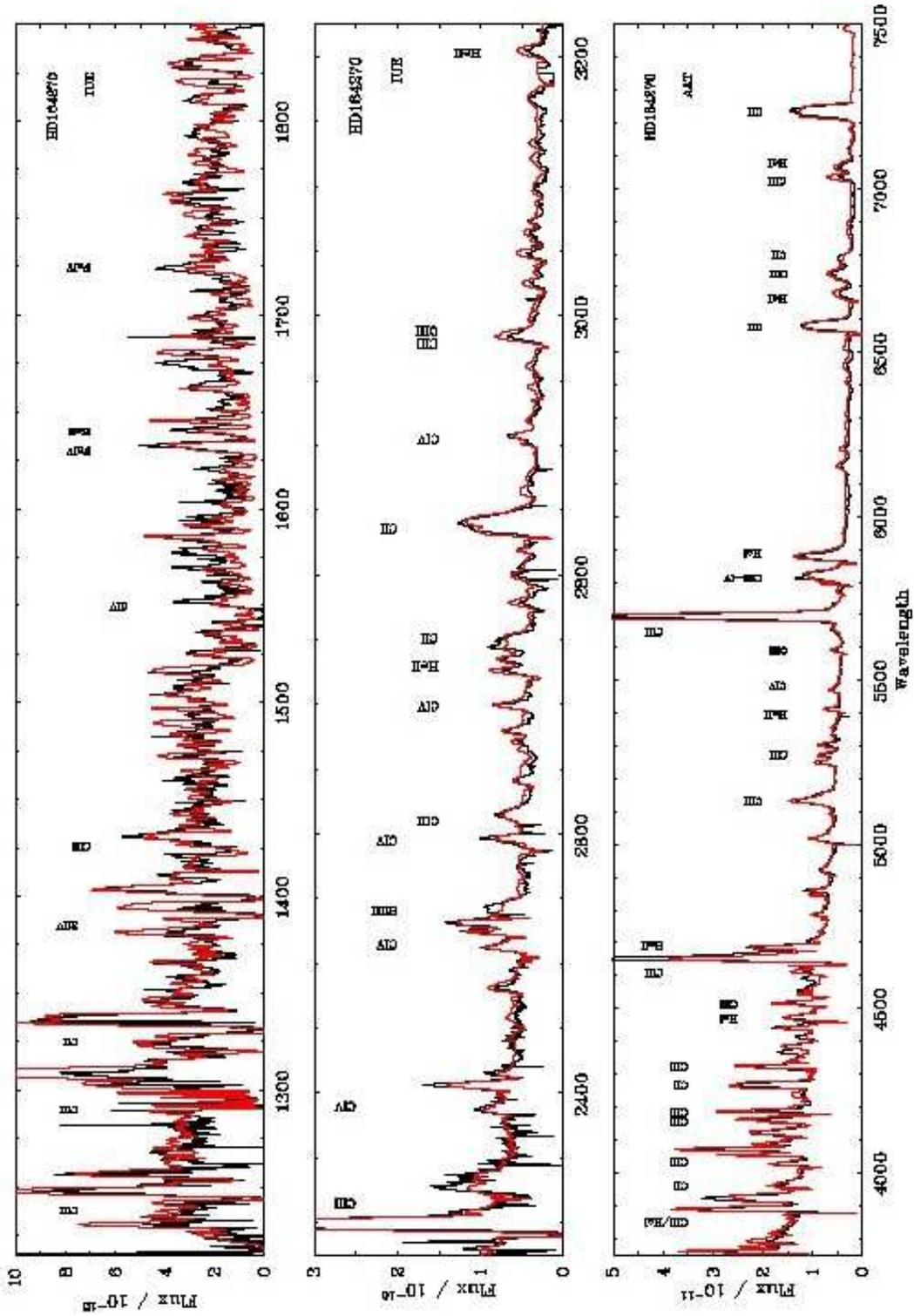}
\end{center}
\caption{Spectroscopic fit (red) to de-reddened (E(B-V)=0.58 mag, R=3.0, 
$\log N(HI$/cm$^{-2}$)=21.0),
UV (IUE/HIRES, 0.1\AA\ resolution) and 
optical (AAT/RGO, 1.7\AA\ resolution) spectrophotometry of HD~164270 (WC9, 
black), including  principal line identifications.
\label{wr103_sp}}
\end{figure*}

\begin{figure}
\begin{center}
\includegraphics[width=0.7\columnwidth,clip,angle=-90]{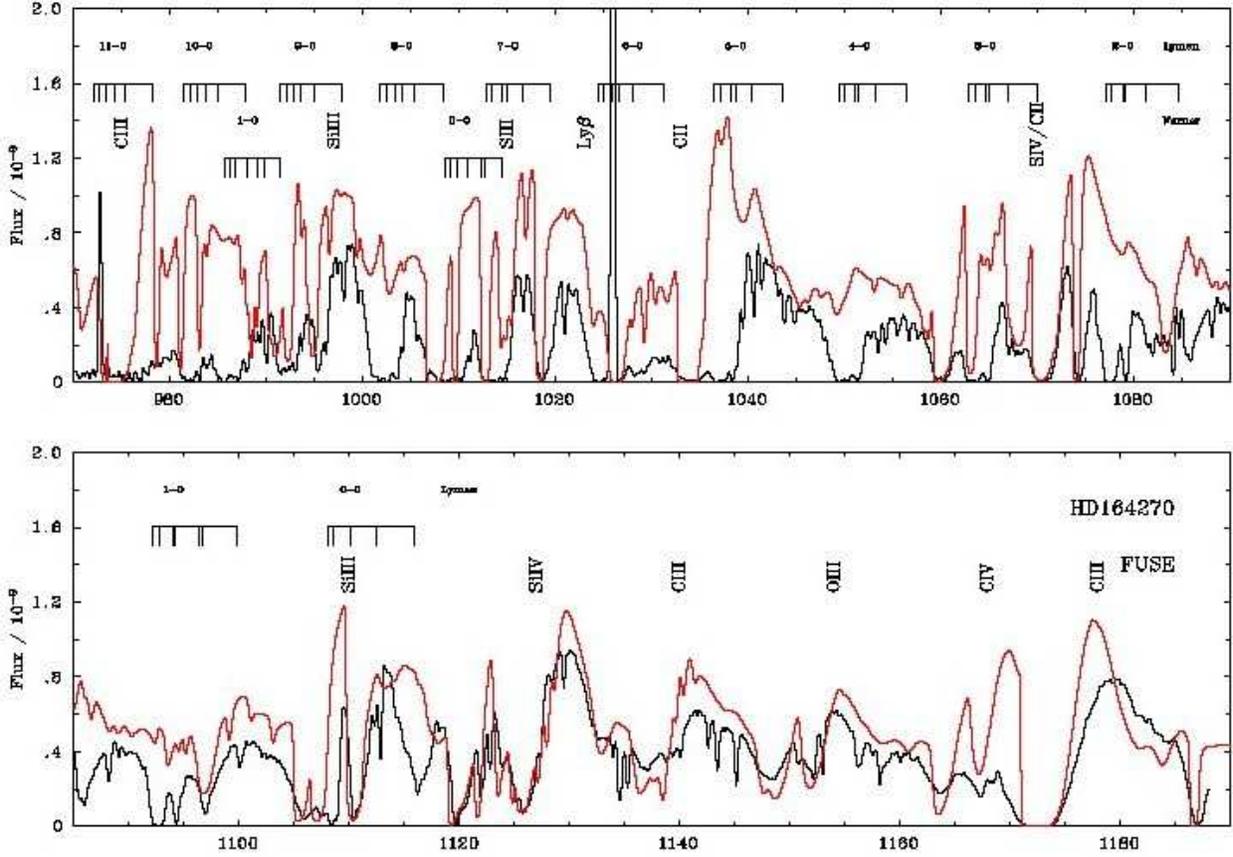}
\end{center}
\caption{Comparison between de-reddened (E(B-V)=0.58 mag, R=3.0) $FUSE$
far-UV  spectrophotometry of HD~164270 (rebinned to 0.1\AA, Willis et al. 
2004)  and our synthetic spectrum (red),  corrected for atomic hydrogen 
($\log N(HI$/cm$^{-2}$)=21.0),
including principal line  identifications. Leading components (J=0-3) of the
Lyman and Werner molecular hydrogen bands are also indicated, which reveal
that the stellar spectral range shortward of $\lambda$1120 is increasingly
contaminated by H$_{2}$.\label{wr103_fuse}}
\end{figure}

\begin{figure*}
\begin{center}
\includegraphics[width=0.85\columnwidth,clip,angle=0]{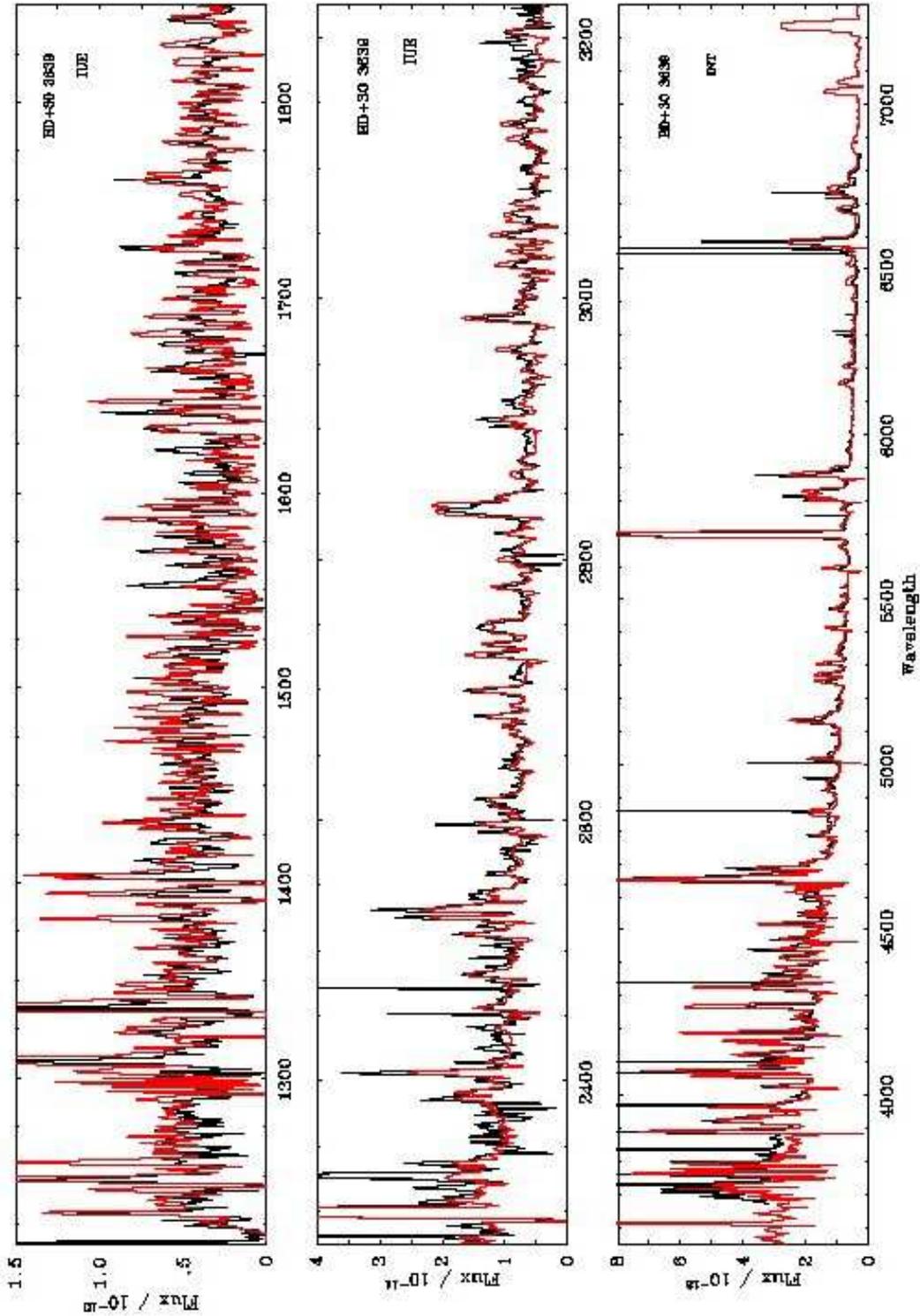}
\end{center}
\caption{Spectroscopic fit (red) to de-reddened (E(B-V)=0.39, R=3.1,
$\log N(HI$/cm$^{-2}$)=21.2), UV (IUE/HIRES, 0.1\AA\ resolution) and 
optical (WHT/ISIS, 3.5\AA\ resolution) spectrophotometry of 
BD+30 3639 ([WC9], black)
\label{bdp30_sp}}
\end{figure*}

\begin{figure}
\begin{center}
\includegraphics[width=0.85\columnwidth,clip]{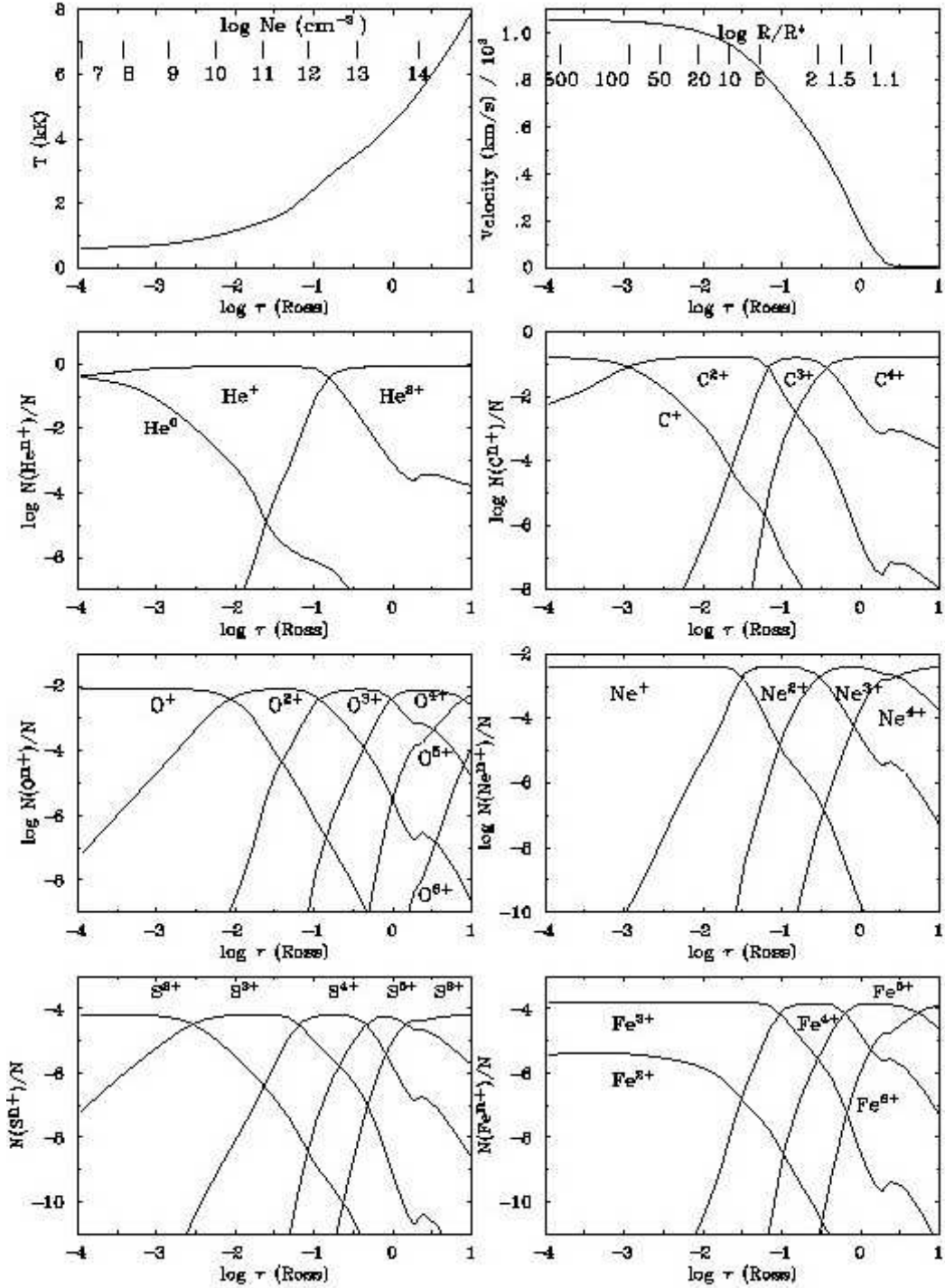}
\end{center}
\caption{Predicted ionization balance of He, C, O, Ne, S and Fe
for our HD~164270 model versus Rosseland optical depth, together with
the electron temperature and density distribution and variation
in wind velocity and radius with optical depth. This figure indicates
that He is partially recombined in the outer wind of WC9 stars, affecting
previous radio derived mass-loss rates (Leitherer et al. 1997; Cappa et al. 2004), 
plus supports dominant ionization  stages of Ne$^{+}$ and 
S$^{2+}$ for neon and sulfur, respectively, in agreement with {\it Spitzer} observations.\label{ion}}
\end{figure}

\begin{figure}
\begin{center}
\includegraphics[width=1.0\columnwidth,clip]{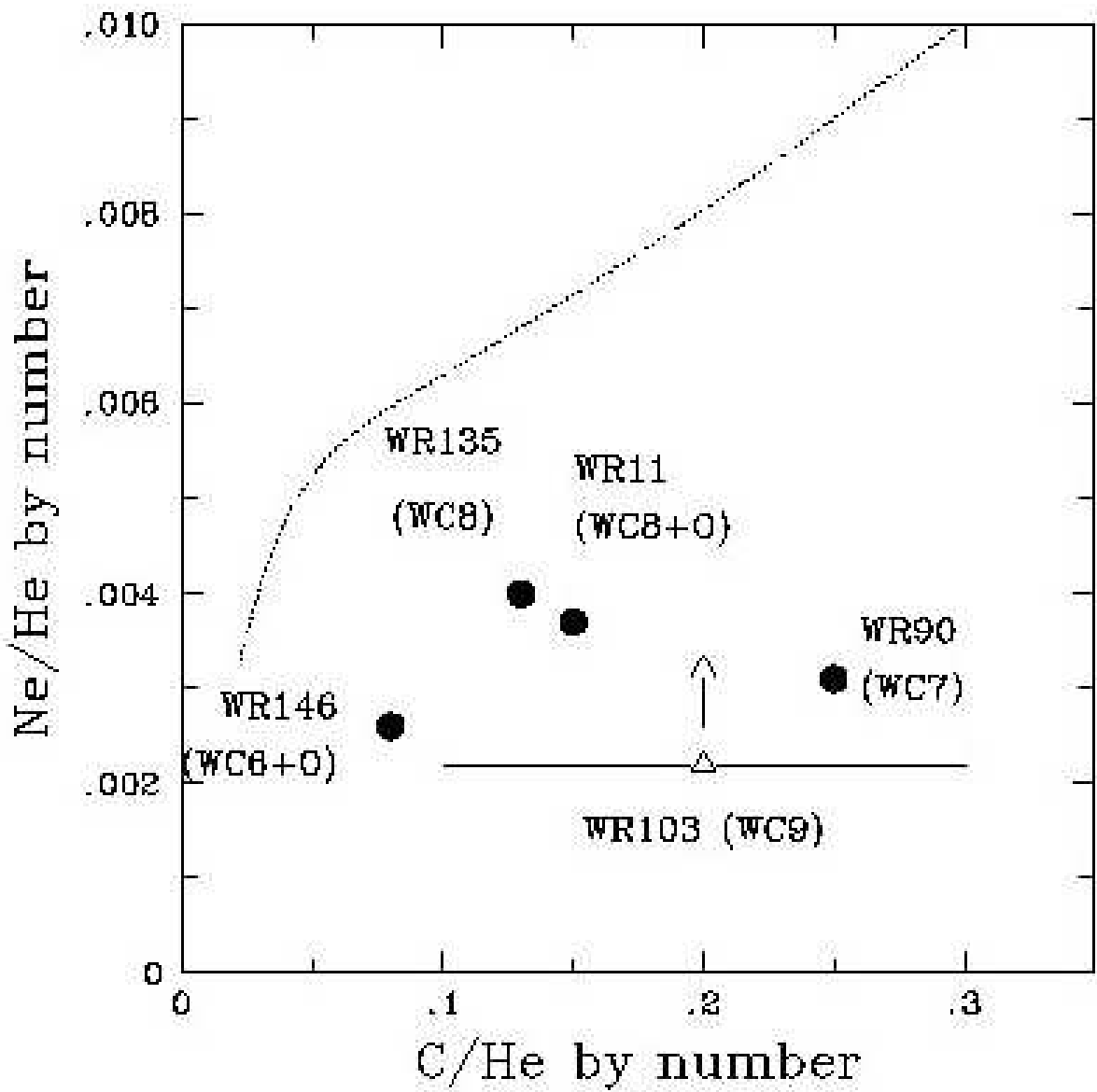}
\end{center}
\caption{Comparison between carbon and neon abundances of 
HD~164270 (WC9, open triangle), by number, relative to other WC stars
analysed in the same manner by Dessart et al. (2000, filled circles).
% Solar C and Ne abundances as derived by Asplund et al. (2004, 2005) are 
% also indicated.
We include predictions for the WC phase of star with  initial mass
60$M_{\odot}$ star at Z=0.02 rotating at 300 km/s from Meynet \& Maeder 
(2003, dotted lines) which indicates that the measured neon abundances
from analysis of mid-IR fine structure lines fall a factor of 2--3 times 
below predictions.
\label{abundances}}
\end{figure}

\begin{figure}
\begin{center}
\includegraphics[width=1.0\columnwidth,clip]{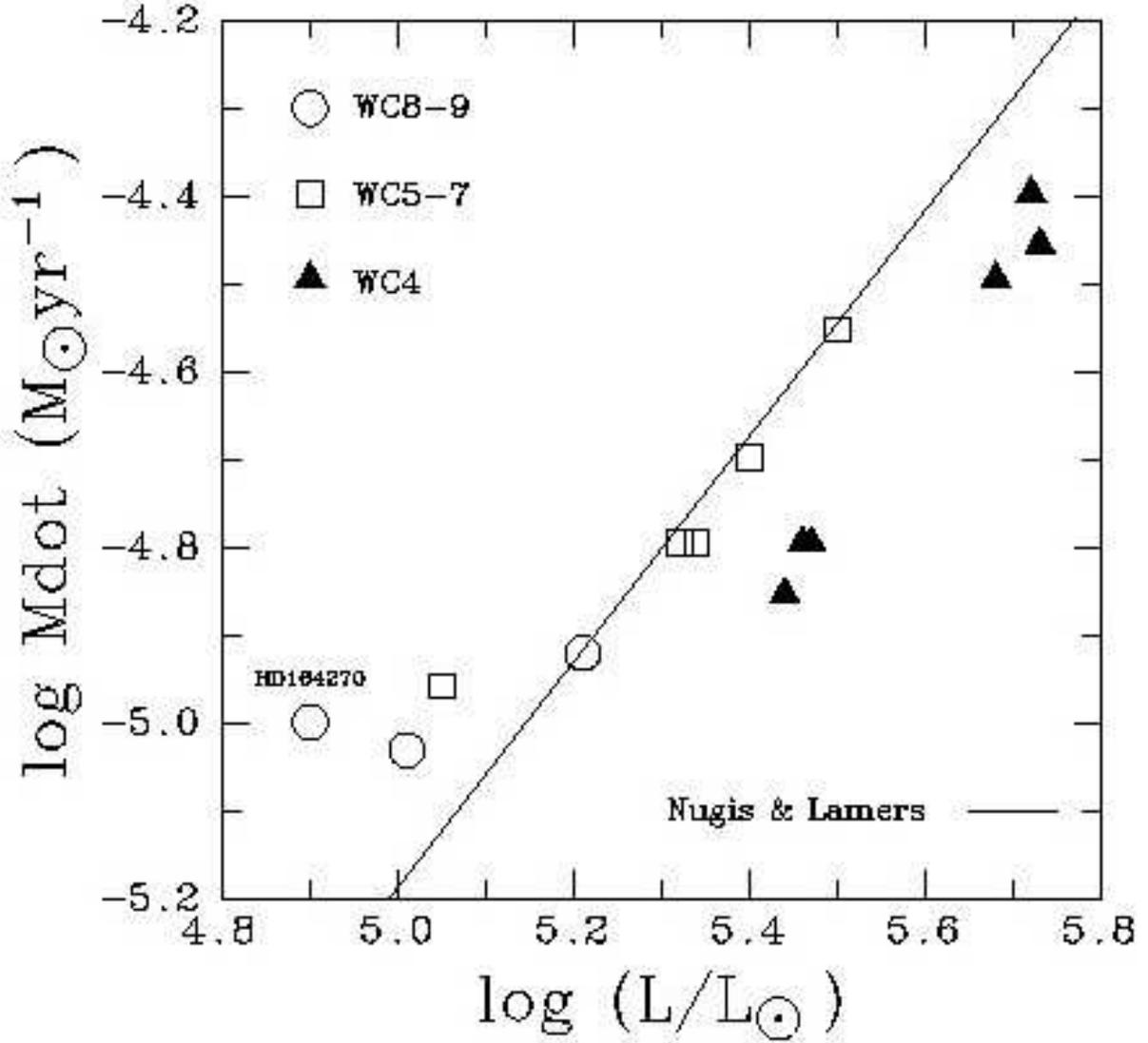}
\end{center}
\caption{Comparison between mass-loss rates and luminosities of Galactic (open) and LMC (filled) WC stars
from Crowther et al. (2002) and references therein, plus HD~164270 (WC9) newly studied here. As a guide,
the generic Wolf-Rayet mass-loss luminosity relation from Nugis \& Lamers (2000, their Eqn.~22) is presented 
for assumed abundances of C/He=0.25 and C/O=5 by number.\label{mdot}}
\end{figure}

\end{document}